\RequirePackage{fixltx2e}
\documentclass[a4paper,fleqn,reqno,english,journal=cmatex,manuscript=article]{revtex4-2}
\usepackage[T1]{fontenc}
\usepackage[latin9]{inputenc}
\usepackage{babel}
\usepackage{array}
\usepackage{float}
\usepackage{textcomp}
\usepackage{multirow}
\usepackage{amsmath}
\usepackage{amsthm}
\usepackage{amssymb}
\usepackage{graphicx}
\usepackage[unicode=true,pdfusetitle,
 bookmarks=true,bookmarksnumbered=false,bookmarksopen=false,
 breaklinks=false,pdfborder={0 0 1},backref=false,colorlinks=false]
 {hyperref}

\makeatletter

\newcommand{\noun}[1]{\textsc{#1}}
\newcommand{\lyxmathsym}[1]{\ifmmode\begingroup\def\b@ld{bold}
  \text{\ifx\math@version\b@ld\bfseries\fi#1}\endgroup\else#1\fi}

\providecommand{\tabularnewline}{\\}
\newcommand{\lyxdot}{.}

\usepackage{lmodern}

\makeatother

\begin{document}
\title{Post-Transition Metal Sn-Based Chalcogenide Perovskites: A Promising
Lead-Free and Transition Metal Alternative for Stable, High-Performance
Photovoltaics}
\author{Surajit Adhikari{*}, Sankhasuvra Das, and Priya Johari}
\email{sa731@snu.edu.in, priya.johari@snu.edu.in}

\affiliation{Department of Physics, School of Natural Sciences, Shiv Nadar Institution
of Eminence, Greater Noida, Gautam Buddha Nagar, Uttar Pradesh 201314,
India.}
\begin{abstract}
Chalcogenide perovskites (CPs) have emerged as promising materials
for optoelectronic applications due to their stability, non-toxicity,
small bandgaps, high absorption coefficients, and defect tolerance.
Although transition metal-based CPs, particularly those incorporating
Zr and Hf, have been well-studied, they often exhibit higher bandgaps,
lower charge carrier mobility, and reduced efficiency compared to
lead-based halide perovskites (HPs). Tin (Sn), a post-transition metal
with a similar oxidation state (+4) as Zr and Hf in ABX$_{3}$ structures
but with different valence characteristics, remains underexplored
in CPs. Given the influence of valence states on material properties,
Sn-based CPs are of great interest. This study employs density functional
theory (DFT), density functional perturbation theory (DFPT), and many-body
perturbation theory (GW and BSE) to investigate a series of distorted
Sn-based CPs (ASnX$_{3}$, A = Ca, Sr, Ba; X = S, Se). Our results
demonstrate that these perovskites are mechanically stable and exhibit
lower direct G$_{0}$W$_{0}$ bandgaps (0.79\textendash 1.50 eV) compared
to their Zr- and Hf-based counterparts. Analysis of carrier-phonon
interactions reveals that the charge-separated polaronic state is
less stable than the bound exciton state in these materials. Additionally,
polaron-assisted charge carrier mobilities for electrons (21.33\textendash 416.02
cm$^{2}$V$^{-1}$s$^{-1}$) and holes (7.02\textendash 260.69 cm$^{2}$V$^{-1}$s$^{-1}$)
are comparable to or higher than those in lead-based HPs and significantly
exceed those of Zr- and Hf-based CPs, owing to reduced carrier-phonon
coupling. The estimated spectroscopic limited maximum efficiency (24.2\%\textendash 31.2\%)\textemdash confirmed
through perovskite solar cell (PSC) simulations using SCAPS-1D software\textemdash indicates
that these materials are promising candidates for photovoltaic applications.
Overall, this study highlights the potential of Sn as a superior alternative
to transition metals in CPs, particularly for photovoltaic applications,
where smaller electronegativity differences lead to reduced bandgaps
and reduced polaronic effects improved charge carrier mobility. These
findings are expected to stimulate further experimental investigation
into Sn-based CPs.
\end{abstract}
\keywords{Chalcogenide perovskites, optoelectronic properties, density functional
theory, Bethe-Salpeter equation, electron-phonon coupling, device
simulation, photovoltaics}
\maketitle

\section{Introduction:}

Chalcogenide perovskites (CPs)\citep{chapter3-11,chapter3-16,chapter3-17,chapter3-18}
have recently emerged as a potential alternative for the high-performing
inorganic-organic halide perovskites (IOHPs)\citep{chapter2-12,chapter3-8,chapter3-9,chapter3-10}.
While IOHPs face significant obstacles for their large-scale industrial
applications owing to the toxicity of lead and instability due to
the presence of organic cations\citep{chapter3-12,chapter1-16}, CPs
show promise because of their vast availability in the earth's crust,
non-toxic nature, and exceptional stability\citep{chapter3-11,chapter3-13,chapter3-14,chapter3-15}.
In addition to these advantages, CPs exhibit favorable characteristics
such as small electronic bandgap, high absorption coefficient, promising
defect tolerance, good charge carrier mobility, and excellent power
conversion efficiency (PCE)\citep{chapter3-16,chapter3-17,chapter3-18,chapter3-19,chapter5-16},
which point toward their potential application in a variety of optoelectronic
devices.

Similar to the typical three-dimensional halide perovskites (HPs),
the chemical formula of chalcogenide perovskites is expressed as ABX$_{3}$\citep{chapter3-3,chapter3-18},
where A and B stand for divalent alkali-earth metal cations (Ca$^{2+}$,
Sr$^{2+}$, Ba$^{2+}$) and tetravalent transition metal cations (Ti$^{4+}$,
Zr$^{4+}$, and Hf$^{4+}$), respectively, and X is typically a chalcogen
anion such as S$^{2-}$ or Se$^{2-}$. Numerous experimental and theoretical
investigations have demonstrated the successful synthesis of chalcogenide
perovskites and unveiled their intriguing properties\citep{chapter3-11,chapter3-13,chapter3-14,chapter3-15,chapter3-16,chapter3-17,chapter3-18,chapter3-19,chapter3-37,chapter3-38}.
For example, Lelieveld et al.\citep{chapter3-37} had synthesized
the distorted phase\noun{ }of CaZrS$_{3}$, SrZrS$_{3}$, BaZrS$_{3}$,
CaHfS$_{3}$, SrHfS$_{3}$, and BaHfS$_{3}$ in 1980, while later,
the needle-like phase\citep{chapter3-38,chapter3-5,chapter3-23} of
SrZrS$_{3}$, SrZrSe$_{3}$, and SrHfSe$_{3}$ CPs were also synthesized
by various groups in the end of $20^{th}$ century and beginning of
$21^{st}$ century. In 2016, Perera et al.\citep{chapter5-4} used
high-temperature sulfurization of the oxides with CS$_{2}$ to synthesize
AZrS$_{3}$ (A = Ba, Ca, and Sr). On the other hand, in 2015, Sun
et al.\citep{chapter3-18} theoretically affirmed the formation of
two distinct phases of CPs at room temperature: the needle-like phase
(NH$_{4}$CdCl$_{3}$$-$type) and the distorted phase (GdFeO$_{3}$$-$type),
both sharing the orthorhombic structure with same space group $Pnma$
(No. 62), and predicted them to be promising for application in solar
cells. In addition, several first-principles DFT-based studies have
been conducted, indicating that CPs exhibit interesting electronic
and optical properties suitable for use in photovoltaic applications\citep{chapter3-19,chapter1-63,chapter3-3,chapter3-18,chapter5-16}.

Several studies have primarily focused on Zr ({[}Kr{]}4d$^{2}$5s$^{2}$)
and Hf ({[}Xe{]}4f$^{14}$5d$^{2}$6s$^{2}$) transition metal-based
CPs, where the valence band configurations involve $d$- and $s$-
orbitals\citep{chapter3-19,chapter1-63,chapter5-16}. These materials
are demonstrated to exhibit exciton binding energies ranging from
0.02 to 0.26 eV, which are comparable to or higher than those observed
in conventional lead-based HPs (0.01$-$0.10 eV)\citep{chapter5-11,chapter5-12,chapter5-14}.
However, the main drawback of these CPs lies in their reduced charge
carrier mobility (6.84$-$77.59 cm$^{2}$V$^{-1}$s$^{-1}$) and lower
PCE (10.56\%$-$25.02\%) compared to conventional HPs, which have
charge carrier mobility and PCE in the range of 57$-$290 cm$^{2}$V$^{-1}$s$^{-1}$\citep{chapter2-20,chapter5-15}
and 21.15\%$-$28.97\%\citep{chapter5-13,chapter5-17}, respectively.
These limitations are majorly due to prominent polaronic effects and
relatively higher bandgaps, which hinder their effectiveness for solar
cell applications. This makes exploration of alternatives for the
B-site cation even more crucial in order to achieve high polaronic
charge carrier mobility along with high PCE, thereby enhancing the
effectiveness of CPs for solar cell applications.

The polaronic effects are more pronounced in transition metal-based
CPs. However, if the post-transition metal Sn ({[}Kr{]}4d$^{10}$5s$^{2}$5p$^{2}$),
which also exhibits a $+4$ oxidation state, is used in place of transition
metals like Zr and Hf, the polaronic effects could be mitigated due
to different valence states. Additionally, a lower difference in the
electronegativity of Sn and X (S, Se) may offer the potential for
achieving appropriate bandgaps for photovoltaic applications. Recent
first-principles calculations highlight the potential of Sn as a B-site
cation \textit{via}. distorted SrSnX$_{3}$ (X = S, Se) perovskites
as promising photovoltaic materials solely based on their electronic
and optical properties up to the HSE06 level\citep{chapter5-2,chapter5-3}.
Experimental investigations have substantiated this potential by fabricating
distorted CaSnS$_{3}$ perovskite at 500°C\citep{chapter5-1}. Basera
et al.\citep{chapter3-19} have furthermore provided valuable insights
into the photovoltaic properties of distorted CaSnS$_{3}$, as well
as needle-like phases of BaSnS$_{3}$ and SrSnS$_{3}$ perovskites.
However, the challenge posed by these needle-like CPs lies in their
higher bandgap (1.91$-$2.04 eV) and lower theoretical efficiency
(21.80\%). In contrast, the bandgap of distorted CaSnS$_{3}$ falls
within the optimal range at 1.43 eV, with theoretical efficiency reaching
up to 32.45\%\citep{chapter3-19}. Motivated by these findings, we
aim to conduct a comprehensive theoretical analysis of post-transition
metal Sn-based distorted CPs to elucidate their suitable photovoltaic
application features.

Henceforth, we have endeavored to meticulously explore the optoelectronic,
transport, excitonic, and polaronic properties of distorted Sn-based,
ASnX$_{3}$ (A = Ca, Sr, Ba; X = S, Se) CPs utilizing a combination
of density functional theory (DFT)\citep{chapter2-36,chapter2-37},
density functional perturbation theory (DFPT)\citep{chapter1-60},
and advanced many-body perturbation theory based methodologies like
GW and BSE\citep{chapter3-1,chapter3-2}. Initially, the crystal structures
are optimized using semilocal PBE\citep{chapter1-34} exchange$-$correlation
(xc) functional, and their stability has been examined. The electronic
properties of the relaxed structures are calculated using HSE06\citep{chapter1-35}
xc functional as well as G$_{0}$W$_{0}$@PBE\citep{chapter1-69,chapter1-70}
methodology. The investigated CPs exhibit direct bandgaps ranging
from 0.79 to 1.50 eV, which fall within the similar range to that
of conventional lead-based HPs\citep{chapter5-11,chapter5-13} and
are lower than the bandgaps observed in Zr- and Hf-based CPs\citep{chapter3-19,chapter1-63,chapter5-16}.
Following this, optical properties and exciton binding energies ($E_{B}$)
are determined through the Bethe-Salpeter equation (BSE)\citep{chapter1-67,chapter1-68}
method. Furthermore, employing the DFPT technique, the ionic contribution
to the dielectric function is computed, and the Fr\"ohlich model is
invoked to evaluate polaronic attributes such as carrier-phonon coupling
strength and polaron mobility (21.33$-$416.02 cm$^{2}$V$^{-1}$s$^{-1}$).
Conclusively, the estimation of the spectroscopic limited maximum
efficiency (SLME)\citep{chapter3-30} between 24.20\%$-$31.20\% utilizing
the quasiparticle (QP) bandgap and absorption coefficient underscores
ASnX$_{3}$ as a material with significant potential for photovoltaic
applications. This has been further confirmed by performing the conventional
device (FTO/TiO$_{2}$/ASnX$_{3}$/Spiro-OMeTAD/Au) simulations using
SCAPS-1D software \citep{chapter5-7,chapter5-8,chapter5-9,chapter5-10}.

\section{Computational Methodologies and Numerical Simulations:}

\subsection{Computational Details:}

In this work, the state-of-the-art first-principles calculations based
on density functional theory (DFT)\citep{chapter2-36,chapter2-37},
density functional perturbation theory (DFPT)\citep{chapter1-60},
and many-body perturbation theory (MBPT)\citep{chapter3-1,chapter3-2}
were performed using the Vienna Ab initio Simulation Package (VASP)\citep{chapter1-31,chapter1-32}.
In all constituent elements, the valence electrons and atomic core
interactions were described using the projector-augmented wave (PAW)
pseudopotentials\citep{chapter1-33}. The PAW pseudopotentials with
valence-electron configurations considered for Ca, Sr, Ba, Sn, S,
and Se were 3s$^{2}$3p$^{6}$4s$^{2}$, 4s$^{2}$4p$^{6}$5s$^{2}$,
5s$^{2}$5p$^{6}$6s$^{2}$, 4d$^{10}$5s$^{2}$5p$^{2}$, 3s$^{2}$3p$^{4}$,
and 4s$^{2}$4p$^{4}$, respectively. For the structural optimization,
the generalized gradient approximation (GGA)\citep{chapter1-34} based
exchange-correlation (xc) functional of Perdew, Burke, and Ernzerhof
(PBE) was employed, which takes into account the electron-electron
interactions. The plane-wave cutoff energy was set to 400 eV, and
the electronic self-consistent-field iteration energy convergence
threshold was chosen as $10^{-6}$ eV. The lattice constants and coordinates
of all the atoms were fully optimized until the Helmann-Feynman forces
on each atom were less than 0.01 eV/$\textrm{\AA}$. The $\Gamma$-centered $7\times7\times5$
$\mathbf{k}$-point sampling was used for Brillouin zone integration
in order to determine the optimized structures. Visualization for
Electronic and STructural Analysis (VESTA)\citep{chapter2-3} software
package was used to display the optimized crystal structures.

The phonon spectra were calculated using the DFPT method as implemented
in the PHONOPY\citep{chapter3-6} package by considering a $2\times2\times2$
supercell. Since, GGA is known to underestimate the bandgap, the electronic
band structures were computed using the hybrid HSE06\citep{chapter1-35}
xc functional as well as many-body perturbation theory (MBPT) based
GW\citep{chapter1-69,chapter1-70} (G$_{0}$W$_{0}$@PBE) method.
Note that the spin-orbit coupling (SOC) effect was not considered
as it does not impact the bandgap (for details, see the Supplemental
Material) of the considered systems. The effective mass was computed
by SUMO\citep{chapter2-10} using a parabolic fitting of the band
edges. We also carried out Bethe-Salpeter equation (BSE)\citep{chapter1-67,chapter1-68}
based calculations on top of the single-shot GW(G$_{0}$W$_{0}$)@PBE
to precisely estimate the optical properties, which takes explicitly
into account the electron-hole interaction. Here, a $\Gamma$-centered
$3\times3\times2$ $\mathbf{k}$-grid and a converged 640 NBANDS were
used for the GW-BSE calculations. The polarizability calculations
were performed using a frequency grid of 60 points to ensure accurate
results. In our GW calculations, we used the plasmon-pole approximation
(PPA) to model the frequency dependence of the dielectric function
(ALGO = GW0). Additionally, the plane-wave cutoff for the response
function was set to ENCUTGW = 267 eV (2/3 of ENCUT) for both GW and
BSE calculations. The electron-hole kernel for the BSE calculations
was generated by considering 24 occupied and 24 unoccupied bands.
The VASPKIT\citep{chapter1-48} package was used to post-process the
elastic and optical properties. The ionic contribution to the dielectric
constant was also calculated using the DFPT method.

Using the hydrogenic Wannier\textminus Mott (WM)\citep{chapter1-63,chapter2-38}
model, the exciton binding energy ($E_{B}$) for a screened Coulomb
interacting $e-h$ pair is calculated as follows:

\begin{equation}
\mathrm{E_{B}=\left(\frac{\mu^{*}}{m_{0}\varepsilon_{eff}^{2}}\right)R_{\infty}},\label{eq:1}
\end{equation}

where, $\mu^{*}$ represents the reduced mass of the charge carriers,
$m_{0}$ denotes the rest mass of electron, $\mathrm{\varepsilon_{eff}}$
is the effective dielectric constant, and $R_{\infty}$ is the Rydberg
constant.

The phonon screening correction to the exciton binding energy ($E_{B}$)
is given by\citep{chapter3-39}:

\begin{equation}
\Delta E_{B}^{ph}=-2\omega_{LO}\left(1-\frac{\varepsilon_{\infty}}{\varepsilon_{static}}\right)\frac{\left(\sqrt{1+\omega_{LO}/E_{B}}+3\right)}{\left(1+\sqrt{1+\omega_{LO}/E_{B}}\right)^{3}},\label{eq:2}
\end{equation}

where, $\varepsilon_{\infty}$ and $\varepsilon_{static}$ are the
electronic (optical) and static (electronic + ionic) dielectric constants,
and $\omega_{LO}$ is the characteristic phonon angular frequency.
The thermal \textquotedbl B\textquotedbl{} approach of Hellwarth
et al.\citep{chapter2-22} is used to determine $\omega_{LO}$ by
taking the spectral average of the multiple phonon branches (for details,
see the Supplemental Material).

Within the framework of Fr\"ohlich's polaron model, the longitudinal
optical phonons and the electron traveling through the lattice interact
via the dimensionless Fr\"ohlich parameter $\alpha$, which is expressed
as\citep{chapter2-13},

\begin{equation}
\alpha=\frac{1}{4\pi\varepsilon_{0}}\frac{1}{2}\left(\frac{1}{\varepsilon_{\infty}}-\frac{1}{\varepsilon_{static}}\right)\frac{e^{2}}{\hbar\omega_{LO}}\left(\frac{2m^{*}\omega_{LO}}{\hbar}\right)^{1/2},\label{eq:3}
\end{equation}

where $\varepsilon_{0}$ is the permittivity of free space and $m^{*}$
is the carrier effective mass. One can also estimate the polaron energy
($E_{p}$) by knowing the value of $\alpha$ using equation\citep{chapter1-63,chapter3-19}:

\begin{equation}
E_{p}=(-\alpha-0.0123\alpha^{2})\hbar\omega_{LO}\label{eq:4}
\end{equation}

Feynman's extended version of Fr\"ohlich's polaron theory (for a small
$\alpha$) is also used to obtain the effective mass of the polaron
($m_{p}$) as follows\citep{chapter2-23}:

\begin{equation}
m_{p}=m^{*}\left(1+\frac{\alpha}{6}+\frac{\alpha^{2}}{40}+...\right)\label{eq:5}
\end{equation}

Finally, using the Hellwarth polaron model\citep{chapter2-22}, the
polaron mobility is defined as follows:

\begin{equation}
\mu_{p}=\frac{\left(3\sqrt{\pi}e\right)}{2\pi c\omega_{LO}m^{*}\alpha}\frac{\sinh(\beta/2)}{\beta^{5/2}}\frac{w^{3}}{v^{3}}\frac{1}{K(a,b)}\label{eq:6}
\end{equation}

where $e$ is the charge of the electron, $\beta=hc\omega_{LO}/k_{B}T$,
and $w$ and $v$ are the temperature-dependent variational parameters,
and $K(a,b)$ is a function of $\beta$, $w$, and $v$ (for details,
see the Supplemental Material).

\subsection{SCAPS-1D Numerical Simulations:}

SCAPS-1D software is utilized to conduct numerical device simulations
on solar cells to assess their performance and aid in design optimization.
The tool was developed by Prof. M. Burgelman at the Department of
Electronics and Information Systems (ELIS) of the University of Gent,
Belgium\citep{chapter5-5,chapter5-6}. SCAPS solves Poisson's equation,
which correlates the electrostatic potential to the overall charge
density, along with continuity equations for electrons and holes in
the conduction and valence bands, respectively. It can predict device
characteristics such as current density-voltage curve, efficiency,
energy bands, and other properties of the solar cell structure under
illumination. Figure \ref{fig:5}(a) shows the proposed perovskite
solar cell (PSC) structure having the architecture of FTO/TiO$_{2}$/Perovskite/Spiro-OMeTAD/Au.
In this proposed structure, Spiro-OMeTAD was used as the hole transport
layer (HTL), TiO$_{2}$ was used as the electron transport layer (ETL),
FTO (fluorine-doped tin oxide) as the transparent conducting oxide
(TCO), and perovskite as the absorber layer. Gold (Au), having a work
function of 5.1 eV, was employed as the back metallic contact. The
simulation parameters for the HTL, ETL, and FTO, chosen based on theoretical
and experimental results\citep{chapter5-7}, are listed in Table S15
of the Supplemental Material. Table S16 also shows the input parameters
for the perovskite layer, estimated through our theoretical calculations
(for details, see the Supplemental Material). The simulations were
conducted under the illumination of AM1.5G at 300K working temperature.
The Poisson's equation and the continuity equations of both the carriers
(electrons and holes) for SCAPS-1D simulation are written as\citep{chapter5-8,chapter5-9},

\begin{equation}
\frac{\partial^{2}\psi}{\partial x^{2}}=-\frac{q}{\varepsilon}[p(x)-n(x)+N_{D}-N_{A}+\rho_{p}\rho_{n}]
\end{equation}

\begin{equation}
-\left(\frac{1}{q}\right)\frac{\partial J_{p}}{\partial x}+G_{op}-R(x)=\frac{\partial p}{\partial t}
\end{equation}

\begin{equation}
\left(\frac{1}{q}\right)\frac{\partial J_{n}}{\partial x}+G_{op}-R(x)=\frac{\partial n}{\partial t}
\end{equation}

Here $\psi$ is the electrostatic potential, $\varepsilon$ is the
permittivity of the material, $p$ and $n$ are electron and hole
concentrations, $N_{D}$ and $N_{A}$ are donor and acceptor densities,
$J_{n}$ and $J_{p}$ are electron and hole current densities, $\rho_{n}$
and $\rho_{p}$ are electron and hole distribution, $R$ is the recombination,
and $G_{op}$ is the optical generation rate.

The open circuit voltage ($V_{oc}$) of the perovskite solar cell
(PSC) is given by\citep{chapter5-7},

\begin{equation}
V_{oc}=\frac{nk_{B}T}{q}ln(\frac{J_{sc}}{J_{s}}+1)
\end{equation}
where $J_{sc}$ is the short circuit current density and $J_{s}$
is the reverse saturation current. The Fill Factor (FF) is given by\citep{chapter5-10}:

\begin{equation}
FF=\frac{P_{max}}{J_{sc}V_{oc}}
\end{equation}

and the Power Conversion efficiency, PCE ($\eta$) is defined as\citep{chapter5-10}:

\begin{equation}
\eta=\frac{P_{max}}{P_{in}}=\frac{FF\times J_{sc}\times V_{oc}}{P_{in}}
\end{equation}

where $P_{max}$ is the maximum power of the solar cell, and $P_{in}$
is the input solar power equivalent to the AM1.5G Sun spectrum.

\section{Results and Discussions:}

In this study, we undertook a systematic and thorough investigation
into the distorted phases of post-transition metal Sn-based chalcogenide
perovskites ASnX$_{3}$ (A = Ca, Sr, Ba; X = S, Se) to explore their
potential optoelectronic features. The subsequent sections delve into
the stability as well as the structural, electronic properties, transport
phenomena, optical properties, excitonic dynamics, polaronic effects,
and the spectroscopic limited maximum efficiency (SLME) of ASnX$_{3}$
chalcogenide perovskites. This detailed examination and discussion
aims to establish a foundational understanding and provide insights
to guide future experimental endeavors.

\subsection{Structural Properties:}

\begin{figure}[H]
\begin{centering}
\includegraphics[width=1\textwidth,height=1\textheight,keepaspectratio]{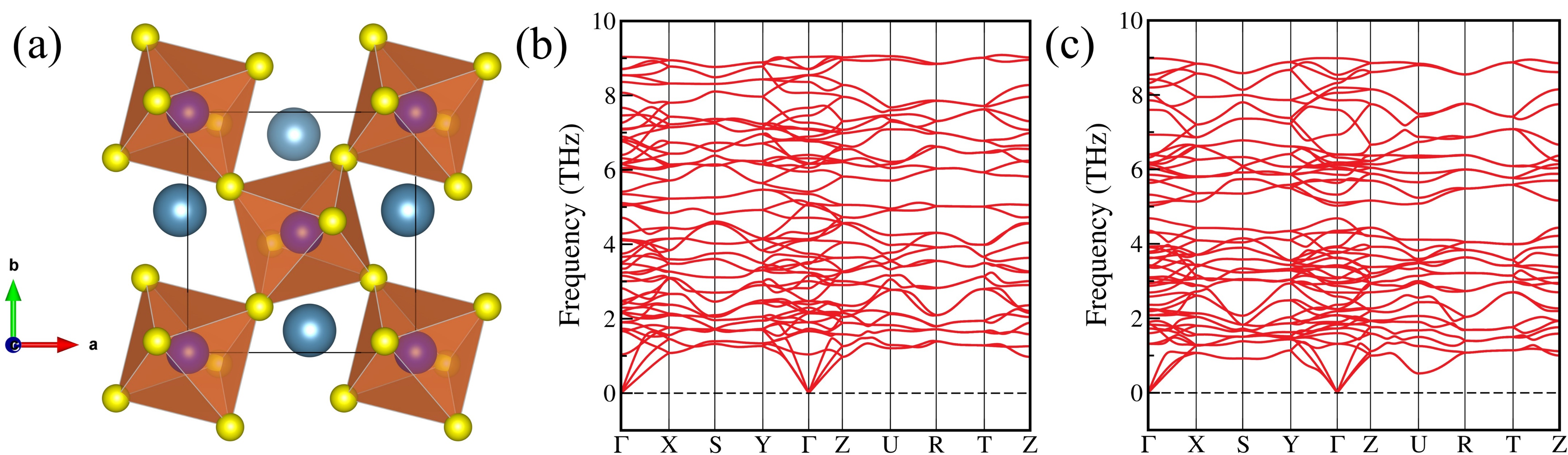}
\par\end{centering}
\caption{\label{fig:1}(a) Crystal structure of CaSnS$_{3}$ in orthorhombic
distorted phase, and phonon dispersion curves of (b) SrSnS$_{3}$,
and (c) BaSnS$_{3}$ calculated using DFPT method. Blue, purple, and
yellow balls represent Ca, Sn, and S atoms, respectively.}
\end{figure}

The orthorhombic distorted crystal structure with the space group
$Pnma$ (No. 62)\citep{chapter3-37} of chalcogenide perovskites ASnX$_{3}$
(A = Ca, Sr, Ba; X = S, Se) are presented in Figure \ref{fig:1}(a)
and Figure S1. The crystal structure of these compounds usually consists
of four formula units, i.e., 20 atoms, of which 4 are Ca/Sr/Ba, 4
are Sn, and 12 are S or Se atoms. In this distorted phase, the alkaline
earth elements (A-site cations) have 12-fold coordination, and they
form cuboctahedrons with chalcogenides X (S or Se). On the other hand,
the corner-sharing distorted octahedrons $[\mathrm{SnX_{6}}]^{8-}$
are produced due to the 6-fold coordination of Sn cation with X (S
or Se) atoms in a distorted and tilted way. The lattice parameters
of the optimized crystal structures are calculated using the PBE xc
functional and are tabulated in Table \ref{tab:1}. It is found that
the lattice parameters are in good agreement with previous theoretical
and available experimental results\citep{chapter5-1,chapter5-2}.
However, for CaSnS$_{3}$, there is a discrepancy in the c-lattice
constant between the theoretical and experimental values, which could
influence the bandgap, as variations in bond lengths and angles modify
the electronic band dispersion. In addition, Table S1 presents detailed
information on the octahedral distortions of SnX$_{6}$ octahedra
in ASnX$_{3}$ (A = Ca, Sr, Ba; X = S, Se) CPs, including parameters
such as average bond length, bond angle variance, polyhedral volume,
and distortion index. The results indicate that the SnX$_{6}$ octahedra
in BaSnX$_{3}$ (X = S, Se) undergo greater lattice distortion than
those in other compounds, likely due to the larger ionic radius of
Ba$^{2+}$, which influences the local bonding environment around
the octahedra.

The crystallographic stability of these CPs is quantitatively identified
by calculating the Goldschmidt tolerance factor ($t$), the octahedral
factor ($\mu$), and the new tolerance factor ($\tau$)\citep{chapter1-37,chapter1-39,chapter2-38}
(for details, see the Supplemental Material). From Table S2, it can
be noticed that the $t$, $\mu$, and $\tau$ values of these investigated
CPs are in the range of 0.879$-$0.964, 0.348$-$0.375, and 4.174$-$4.721,
respectively, confirming the reasonable structural stability of these
distorted CPs\citep{chapter1-37,chapter3-3}. Furthermore, the decomposition
energy is also calculated for these materials using the PBE xc functional
to ascertain their thermodynamic stability (for details, see Table
S3 of the Supplemental Material). The negative decomposition energy
reveals these perovskites to be unstable at 0 K. Still, there remains
a possibility for them to stabilize at higher temperatures, as also
demonstrated by Hamza et al. by synthesizing CaSnS$_{3}$ perovskite
at 500$\lyxmathsym{\textdegree}$ C\citep{chapter5-1}, which is an
indicator of their higher temperature stability. Further, the dynamical
and mechanical stability of the studied CPs are examined.

\begin{table}[H]
\caption{\label{tab:1}Calculated lattice parameters of ASnX$_{3}$ (A = Ca,
Sr, Ba; X = S, Se) chalcogenide perovskites.}

\centering{}%
\begin{tabular}{ccccccccccc}
\hline 
\multirow{2}{*}{Configurations} &  & \multicolumn{3}{c}{This Work} &  &  & \multicolumn{3}{c}{Previous Work} & \multirow{2}{*}{Reference}\tabularnewline
\cline{3-5} \cline{4-5} \cline{5-5} \cline{8-10} \cline{9-10} \cline{10-10} 
 &  & a ($\textrm{\AA}$) & b ($\textrm{\AA}$) & c ($\textrm{\AA}$) &  &  & a ($\textrm{\AA}$) & b ($\textrm{\AA}$) & c ($\textrm{\AA}$) & \tabularnewline
\hline 
\multirow{1}{*}{CaSnS$_{3}$} &  & 6.71 & 7.08 & 9.67 &  &  & 6.69 & 7.08 & 11.29 & Expt.\citep{chapter5-1}\tabularnewline
\multirow{1}{*}{SrSnS$_{3}$} &  & 6.90 & 7.22 & 9.82 &  &  & 6.90 & 7.21 & 9.84 & Theo.\citep{chapter5-2}\tabularnewline
BaSnS$_{3}$ &  & 7.04 & 7.27 & 10.17 &  &  &  &  &  & \tabularnewline
CaSnSe$_{3}$ &  & 7.06 & 7.47 & 10.15 &  &  &  &  &  & \tabularnewline
SrSnSe$_{3}$ &  & 7.24 & 7.62 & 10.29 &  &  & 7.25 & 7.61 & 10.31 & Theo.\citep{chapter5-2}\tabularnewline
BaSnSe$_{3}$ &  & 7.32 & 7.73 & 10.63 &  &  &  &  &  & \tabularnewline
\hline 
\end{tabular}
\end{table}

In Figure \ref{fig:1}(b)-(c), the phonon spectra of SrSnS$_{3}$
and BaSnS$_{3}$ CPs have been depicted, which validates their dynamical
stability at 0 K. The rest of the compounds are found to be unstable
at 0 K, but they may become dynamically stable at higher temperatures,
as discussed before. To investigate the mechanical stability, the
second-order elastic coefficients ($C_{ij}$) of these CPs are calculated
using the energy-strain approach\citep{chapter1-47} (for details,
see the Supplemental Material). The computed $C_{ij}$ values of these
compounds are listed in Table S4, and they are found to satisfy the
Born stability criteria\citep{chapter1-47}. Using these elastic coefficients,
the bulk modulus ($B$), shear modulus ($G$), Young\textquoteright s
modulus ($Y$), and Poisson\textquoteright s ratio ($\nu$) of these
materials\citep{chapter1-49,chapter1-50} are investigated (for details,
see the Supplemental Material). The fragility of the materials is
studied in terms of Pugh\textquoteright s suggested ratio ($B/G$)\citep{chapter1-51}
and Poisson\textquoteright s ratio ($\nu$). The calculated values
of $B/G$ (> 1.75) and $\nu$ (> 0.26) reveal that the examined CPs
are ductile in nature (see Table S4). Additionally, the longitudinal
($v_{l}$), transverse ($v_{t}$), and average ($v_{m}$) elastic
wave velocities and the Debye temperature ($\Theta_{D}$)\citep{chapter3-26}
are calculated (see Table S5), as they hold crucial significance for
flexible optoelectronic applications.

\begin{table}[H]
\caption{\label{tab:2}Bandgap (in eV) of chalcogenide perovskites. Here, Ca-
and Sr-based CPs are direct bandgap materials, whereas Ba-based CPs
are indirect bandgap materials. For Ba-based CPs, $i$ and $d$ represent
indirect and direct bandgaps, respectively.}

\centering{}{\small{}}%
\begin{tabular}{ccccccccccccc}
\hline 
\multicolumn{1}{c}{{\small{}Configurations}} &  &  & {\small{}PBE} &  &  & {\small{}HSE06} &  &  & {\small{}G$_{0}$W$_{0}$@PBE} &  &  & {\small{}Previous Work}\tabularnewline
\hline 
{\small{}CaSnS$_{3}$} &  &  & {\small{}0.77} &  &  & {\small{}1.40} &  &  & {\small{}1.44} &  &  & {\small{}1.72 (Expt.\citep{chapter5-1})}\tabularnewline
{\small{}SrSnS$_{3}$} &  &  & {\small{}0.83} &  &  & {\small{}1.45} &  &  & {\small{}1.50} &  &  & {\small{}1.56 (Theo.\citep{chapter5-3})}\tabularnewline
{\small{}BaSnS$_{3}$} &  &  & {\small{}$0.64^{i}$ ($0.66^{d}$)} &  &  & {\small{}$1.16^{i}$ ($1.23^{d}$)} &  &  & {\small{}$1.18^{i}$ ($1.28^{d}$)} &  &  & \tabularnewline
{\small{}CaSnSe$_{3}$} &  &  & {\small{}0.25} &  &  & {\small{}0.70} &  &  & {\small{}0.79} &  &  & \tabularnewline
{\small{}SrSnSe$_{3}$} &  &  & {\small{}0.42} &  &  & {\small{}0.86} &  &  & {\small{}0.88} &  &  & {\small{}1.00 (Theo.\citep{chapter5-3})}\tabularnewline
{\small{}BaSnSe$_{3}$} &  &  & {\small{}$0.54^{i}$ ($0.55^{d}$)} &  &  & {\small{}$0.93^{i}$ ($1.00^{d}$)} &  &  & {\small{}$0.99^{i}$ ($1.04^{d}$)} &  &  & \tabularnewline
\hline 
\end{tabular}{\small\par}
\end{table}

\subsection{Electronic Properties:}

\begin{figure}[H]
\begin{centering}
\includegraphics[width=1\textwidth,height=1\textheight,keepaspectratio]{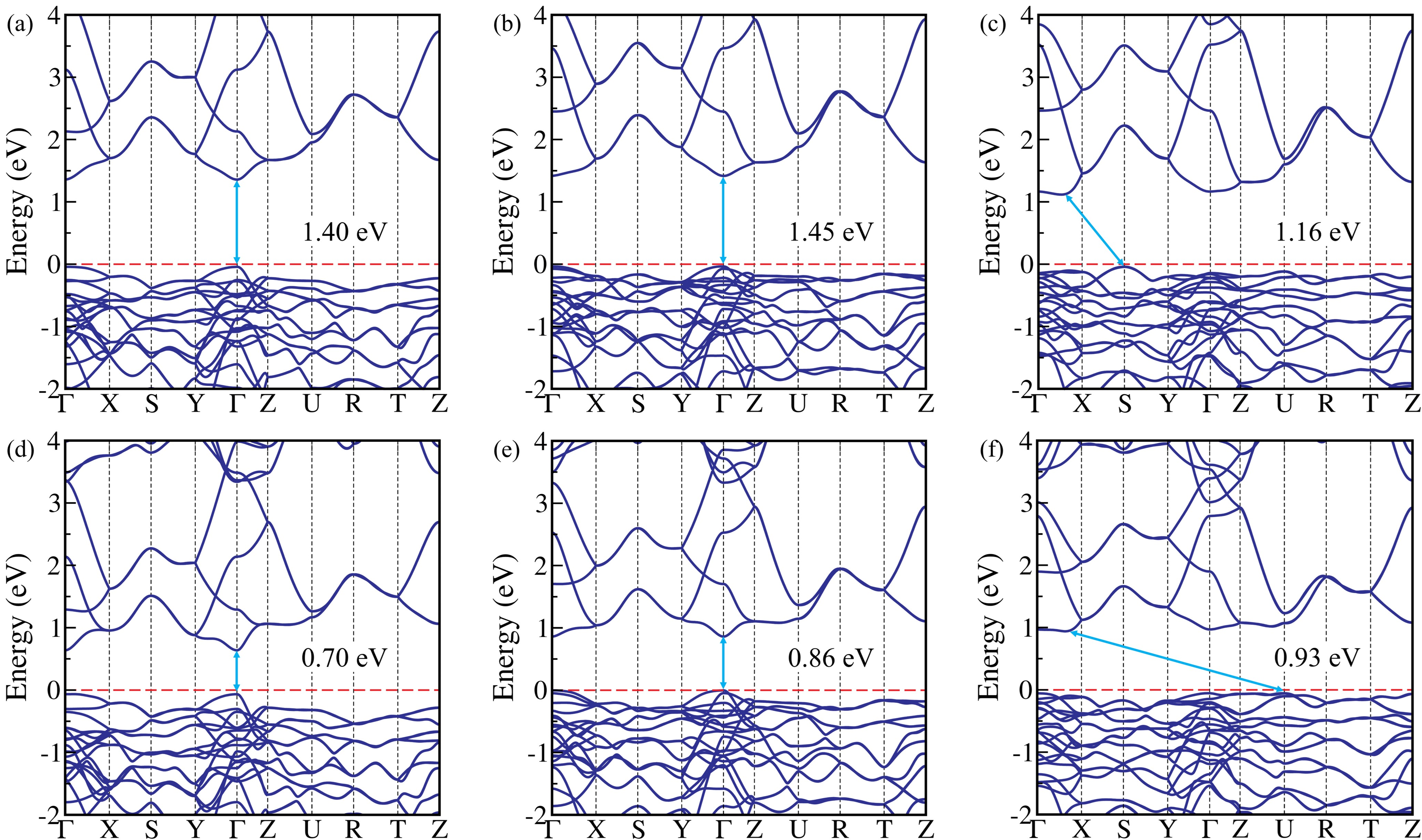}
\par\end{centering}
\caption{\label{fig:2}Electronic band structures of (a) CaSnS$_{3}$, (b)
SrSnS$_{3}$, (c) BaSnS$_{3}$, (d) CaSnSe$_{3}$, (e) SrSnSe$_{3}$,
and (f) BaSnSe$_{3}$ chalcogenide perovskites, obtained using the
HSE06 xc functional. The Fermi level is set to be zero and marked
by the dashed line.}
\end{figure}

After examining the stability, the electronic properties, such as
band structure and partial density of states (PDOS) of the chalcogenide
perovskites ASnX$_{3}$ (A = Ca, Sr, Ba; X = S, Se) are computed to
gain deep insights for designing the photoelectric devices. At first,
the electronic band structure calculations are performed using the
semi-local GGA-PBE xc functional with and without including the spin-orbit
coupling (SOC) for these CPs. It is found that GGA-PBE xc functional
is unable to predict the correct bandgaps due to the self-interaction
error of the electrons (see Table \ref{tab:2}), and SOC does not
have any impact on the bandgap (see Table S6 of the Supplemental Material),
which is expected for chalcogenide perovskites\citep{chapter1-63,chapter3-19}.
After that, we employed the hybrid HSE06 xc functional and many-body
perturbation theory (MBPT) based GW (G$_{0}$W$_{0}$@PBE) method
to calculate the bandgaps more accurately. The HSE06 and G$_{0}$W$_{0}$@PBE
calculated band structures of these compounds are depicted in Figure
\ref{fig:2} and Figure S3, respectively. Our results revealed that
Ca- and Sr-based compounds have direct bandgaps at $\Gamma$ (0, 0,
0) point, while Ba-based CPs exhibit indirect bandgaps. In the case
of BaSnS$_{3}$ and BaSnSe$_{3}$, the valence band maxima (VBM) are
situated at the S (0.5, 0.5, 0) and U (0.5, 0, 0.5) points of the
Brillouin zone (BZ), respectively, while the conduction band minima
(CBM) are found in between $\Gamma$ and X (0.5, 0, 0) point for both
cases (see Figure \ref{fig:2}). Also, both of them exhibit the lowest
direct bandgap in between $\Gamma$ and X point of the BZ. The HSE06
as well as G$_{0}$W$_{0}$@PBE estimated bandgaps of these CPs are
listed in Table \ref{tab:2}, which closely match with earlier theoretical
findings. It is well known that the G$_{0}$W$_{0}$@PBE method provides
highly accurate bandgap values for CPs, yielding results that closely
match experimental values\citep{chapter3-19,chapter1-63,chapter5-16}.
However, the G$_{0}$W$_{0}$@PBE bandgap of CaSnS$_{3}$ (1.44 eV)
is in reasonable agreement with its experimental bandgap of 1.72 eV\citep{chapter5-1},
showing a difference of 0.28 eV. The underestimation of the GW bandgap
may partly arise from structural discrepancies (c-lattice constant),
as GW calculations are sensitive to input geometries, and deviations
from experimental structures can affect electronic properties. Additionally,
the experimental bandgap of CaSnS$_{3}$ was measured at 500°C (773.15
K), while theoretical calculations are performed at 0 K, and this
temperature difference could also contribute to the discrepancy. Furthermore,
incorporating excitonic effects into the calculations could potentially
lead to a larger difference between the theoretical (optical) and
experimental bandgaps. The HSE06 and G$_{0}$W$_{0}$@PBE computed
bandgap of these CPs lie in the range of 0.70$-$1.45 eV and 0.79$-$1.50
eV, respectively. Thus, the bandgaps of the investigated post-transition
metal Sn-based CPs fall within the optimal range for photovoltaic
applications, aligning closely with those of conventional APbI$_{3}$
(A = Cs, CH$_{3}$NH$_{3}$) HPs (1.50$-$1.85 eV)\citep{chapter5-11,chapter5-13}
and being lower than the bandgaps of transition metal based CPs (1.69$-$2.46
eV)\citep{chapter3-19,chapter1-63,chapter5-16} (for details, see
Table S14 of the Supplemental Material). This suggests their high
suitability for photovoltaic applications.

Figure S4 shows the partial density of states (PDOS) and the total
density of states (TDOS) for all the studied compounds obtained using
the HSE06 xc functional. In ASnS$_{3}$ (A = Ca, Sr, Ba) CPs, S-$3p$
orbitals mainly contribute to the VBM, while the hybridization (sp$^{3}$-type)
of S-$3p$ and Sn-$5s$ orbitals dominates the CBM. Similarly, for
ASnSe$_{3}$ perovskites, the VBM is primarily contributed by Se-$4p$
orbitals, whereas CBM mainly consists of hybridized Se-$4p$ and Sn-$5s$
orbitals. The hybridization of Se-$4p$ and Sn-$5s$ orbitals in ASnSe$_{3}$
is stronger than the hybridization of S-$3p$ and Sn-$5s$ orbitals
in ASnS$_{3}$, which is mainly responsible for lowering the bandgap
in the case of ASnSe$_{3}$ as compared to their sulfur counterparts
ASnS$_{3}$. On the other hand, in the case of transition metal-based
CPs, the bonding (sp$^{3}$d$^{2}$) is a mix of $\sigma$$-$bonding
(from the covalent overlap of orbitals) and $\pi$$-$bonding (to
a lesser extent, from $d$-orbital interactions), leading to strong
covalent interactions. At the same time, a more significant difference
in the electronegativity of Zr/Hf and S/Se adds some ionic characteristics
(see Figure S5), which eventually results in more localized electronic
states around the atom and, thus, larger bandgaps than Sn-based CPs\citep{chapter3-19,chapter1-63,chapter5-16}.

Notably, perovskites are highly recognized materials for efficiently
transporting charge carriers (electrons and holes). Therefore, in
addition to their band structures, we also calculate the corresponding
electron ($m_{e}^{*}$) and hole ($m_{h}^{*}$) effective masses by
utilizing the fitted E\textminus k dispersion band diagram of G$_{0}$W$_{0}$@PBE
band structures (Figure S3), employing the formula, $\text{\ensuremath{m^{*}=\hbar^{2}\left[\partial^{2}E(k)/\partial k^{2}\right]^{-1}}}$
(for details, see the Supplemental Material). From Table \ref{tab:3},
it is observed that $m_{e}^{*}$ < 1, $m_{h}^{*}$ < 1.5 for all the
cases, indicating high carrier mobility and thus better charge carrier
transport. Since BaSnX$_{3}$ (X = S, Se) exhibits an indirect bandgap,
the effective masses for these compounds are calculated at both their
indirect and direct band edges. Also, Table \ref{tab:3} suggests
that Ca- and Sr-based CPs have a high potential for exhibiting ambipolar
characteristics.

\begin{table}[H]
\caption{\label{tab:3}Effective mass of electron ($m_{e}^{*}$) and hole ($m_{h}^{*}$)
and their reduced mass ($\mu^{*}$), obtained using the G$_{0}$W$_{0}$@PBE
method. All values are in terms of free-electron mass ($m_{0}$) and
the bold values provided in parentheses are the effective mass and
respective reduced mass at direct band edge.}

\centering{}%
\begin{tabular}{ccccccccccccc}
\hline 
\multicolumn{1}{c}{Configurations} &  &  &  & $m_{e}^{*}$ ($m_{0}$) &  &  &  & $m_{h}^{*}$ ($m_{0}$) &  &  &  & $\mu^{*}$ ($m_{0}$)\tabularnewline
\hline 
CaSnS$_{3}$ &  &  &  & 0.403 &  &  &  & 0.537 &  &  &  & 0.230\tabularnewline
SrSnS$_{3}$ &  &  &  & 0.608 &  &  &  & 0.673 &  &  &  & 0.319\tabularnewline
BaSnS$_{3}$ &  &  &  & 0.674 (\textbf{0.674}) &  &  &  & 1.287 (\textbf{1.121}) &  &  &  & 0.442 (\textbf{0.421})\tabularnewline
CaSnSe$_{3}$ &  &  &  & 0.149 &  &  &  & 0.207 &  &  &  & 0.087\tabularnewline
SrSnSe$_{3}$ &  &  &  & 0.240 &  &  &  & 0.385 &  &  &  & 0.148\tabularnewline
BaSnSe$_{3}$ &  &  &  & 0.717 (\textbf{0.717}) &  &  &  & 1.059 (\textbf{1.237}) &  &  &  & 0.428 (\textbf{0.454})\tabularnewline
\hline 
\end{tabular}
\end{table}

\subsection{Optical Properties:}

In addition to the electronic bandgaps, we also explore the optical
bandgaps to gain a holistic understanding of the excitonic effects
in perovskite materials. These effects, arising from the interaction
of electrons and holes, are pivotal for determining the efficiency
of light-matter interactions in optoelectronic applications. The above
study shows that the electronic properties of these CPs can be well-described
by the HSE06 functional. Nevertheless, this functional is known to
predict the optical features of these systems with less accuracy.
Thus, we performed MBPT-based GW-BSE calculations to calculate the
optical properties. In essence, GW calculations compute the fundamental
bandgap, which is thought to be more accurate and comparable to photoelectron
spectroscopy (PES) and inverse photoelectron spectroscopy (IPES)\citep{chapter1-69,chapter1-70}.
In contrast, BSE calculations predict the optical bandgap similar
to experimental optical absorption spectroscopy\citep{chapter1-67,chapter1-68}.

\begin{figure}[H]
\begin{centering}
\includegraphics[width=1\textwidth,height=1\textheight,keepaspectratio]{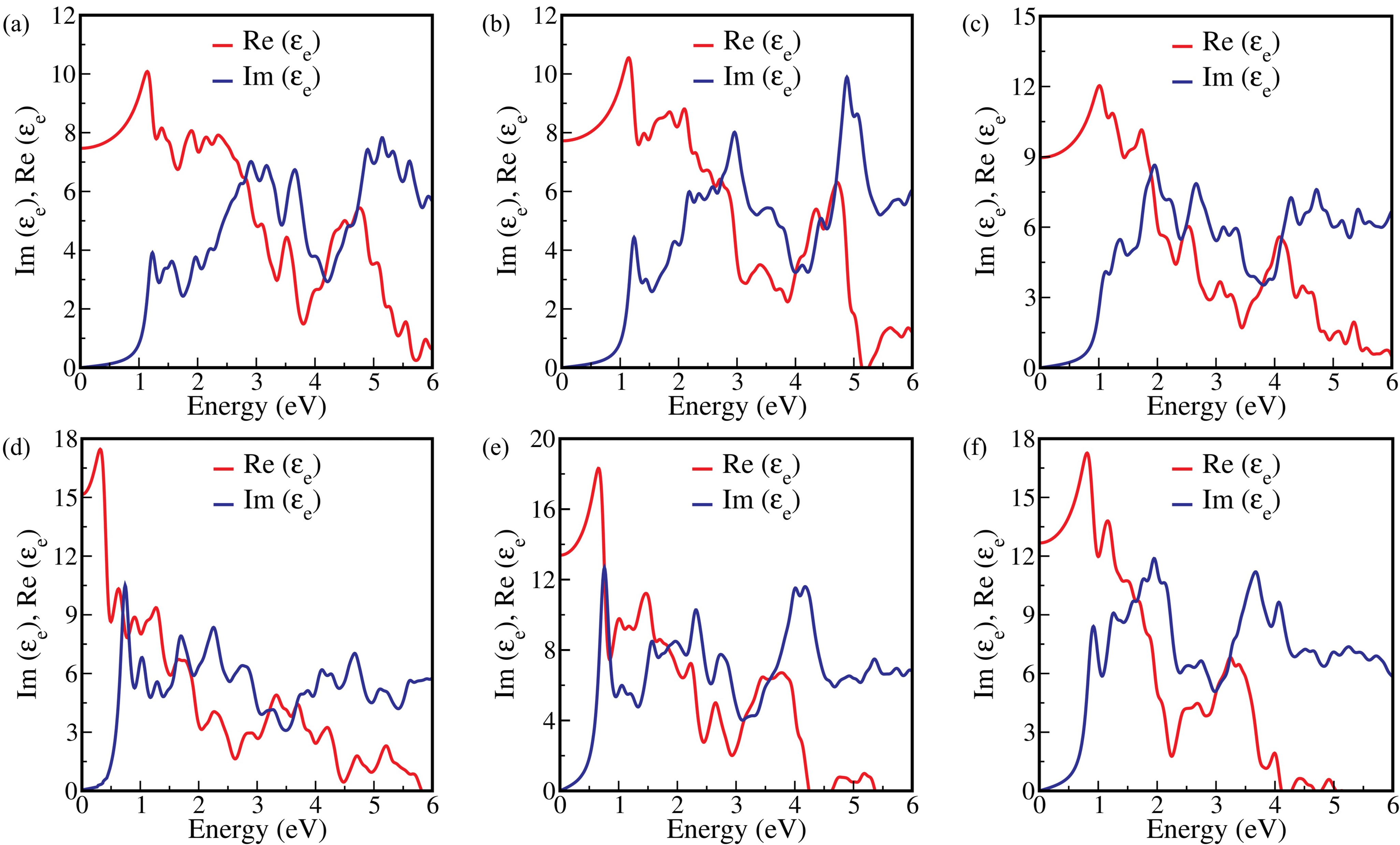}
\par\end{centering}
\caption{\label{fig:3}Computed real {[}Re($\varepsilon_{e}$){]} and imaginary
{[}Im($\varepsilon_{e}$){]} part of the electronic dielectric function
for (a) CaSnS$_{3}$, (b) SrSnS$_{3}$, (c) BaSnS$_{3}$, (d) CaSnSe$_{3}$,
(e) SrSnSe$_{3}$, and (f) BaSnSe$_{3}$ chalcogenide perovskites,
obtained using the BSE@G$_{0}$W$_{0}$@ PBE method.}
\end{figure}

To evaluate the optical responses of the ASnX$_{3}$ CPs, BSE calculations
are performed on top of the single-shot GW(G$_{0}$W$_{0}$)@PBE,
which explicitly considers the electron-hole interaction. The real
{[}Re ($\varepsilon_{e}$){]} and imaginary {[}Im ($\varepsilon_{e}$){]}
part of the frequency-dependent electronic dielectric function calculated
using BSE@G$_{0}$W$_{0}$@ PBE are shown in Figure\ref{fig:3}. It
is discovered that the absorption onset and the first peak position
or optical bandgap ($E_{o}$) gradually red shift from sulfide (S)
to selenide (Se) containing CPs akin to the drop of quasiparticle
(QP) bandgap of them (see Table \ref{tab:2}). For example, the optical
bandgap for CaSnS$_{3}$ is observed at 1.23 eV, and it shifts to
0.74 eV for CaSnSe$_{3}$. After incorporating the excitonic effect
in CaSnS$_{3}$, it is evident that the difference between the theoretical
(optical) bandgap and the experimental bandgap increases by 0.49 eV.
However, the underlying reason for this difference remains consistent
with the previously discussed influence of structural geometry and
temperature. The value of $E_{o}$ for these ASnX$_{3}$ CPs falls
within the range of 0.74 to 1.26 eV and shifts with the change of
the A atom, similar to the variation observed in their QP bandgap
(see Table \ref{tab:4}). Our calculations reveal that the optical
bandgaps of these CPs are consistently smaller than their electronic
bandgaps due to significant excitonic effects. This study primarily
focuses on the bright excitonic states, however, dark excitons can
significantly influence the optical properties of semiconductors,
particularly in systems with low symmetry or strong spin-orbit coupling.
Dark excitons, which are optically inactive due to spin-forbidden
transitions or momentum mismatch, are typically found below the bright
excitons in the absorption spectrum. Although they do not couple directly
to light, their presence can affect the overall photophysical behavior,
such as photoluminescence and carrier dynamics. Nevertheless, an analysis
of the BSE eigenvalues revealed no evidence of dark excitons (optically
inactive states) below the first bright exciton for the investigated
materials.

The electronic dielectric constants ($\varepsilon_{\infty}$), which
are obtained from the real part of the dielectric function, are also
found to be increased from S to Se containing CPs, indicating low
charge carrier recombination rate and improved optoelectronic efficiency
for the latter\citep{chapter2-48} (see Table \ref{tab:4}). For instance,
$\varepsilon_{\infty}$ for CaSnS$_{3}$ is 7.47, and it increases
to 15.15 for CaSnSe$_{3}$. Overall, these results reveal that CaSnS$_{3}$,
SrSnS$_{3}$, BaSnS$_{3}$, and BaSnSe$_{3}$ could be the best choice
for photovoltaic applications.

The absorption coefficient {[}$\alpha(\omega)${]} of a material is
a measurable parameter that indicates how many photons (of a specific
wavelength) may enter the material before being absorbed. Thus, this
descriptor is also essential for photovoltaic applications and offers
crucial information regarding the ideal solar energy conversion efficiency.
$\alpha(\omega)$ is associated with the dielectric function and can
be computed with the following formula\citep{chapter1-60}:
\begin{equation}
\alpha(\omega)=\sqrt{2}\omega\text{\ensuremath{\left[\sqrt{(Re(\varepsilon_{e}))^{2}+(Im(\varepsilon_{e}))^{2}}-Re(\varepsilon_{e})\right]^{1/2}.}}
\end{equation}

Our results from the BSE@G$_{0}$W$_{0}$@ PBE calculation (see Figure
S6) imply that all the investigated ASnX$_{3}$ CPs exhibit high optical
absorption coefficients ($\sim$ 10$^{4}$ $-$ 10$^{5}$ cm$^{-1}$),
which is desirable for the photovoltaic applications.

\subsection{Excitonic Properties:}

In optoelectronic materials, exciton generation plays a significant
role in the charge separation properties. For this reason, excitonic
parameters, such as exciton binding energy ($E_{B}$) and exciton
lifetime ($\tau_{exc}$), are crucial descriptors in their applications.
The energy required to dissociate an exciton into a single electron
($e$) and hole ($h$) pair is known as the exciton binding energy.
The GW-calculated electronic gaps are complemented by BSE-computed
optical gaps to account for excitonic binding energies. This combined
approach allows us to predict material properties with high accuracy,
ensuring their relevance for practical device design. Theoretically,
$E_{B}$ is computed as the difference of the QP bandgap (direct G$_{0}$W$_{0}$@PBE
bandgap, $E_{g}^{d}$) and the optical bandgap (BSE@G$_{0}$W$_{0}$@PBE
peak position, $E_{o}$)\citep{chapter2-16,chapter2-17,chapter2-38,chapter5-18},
which are tabulated in Table \ref{tab:4}. It should be noted that
the electronic contribution of the dielectric screening is dominant
over the ionic one if $E_{B}$ is significantly greater than the longitudinal
optical phonon energy ($\hbar\omega_{LO}$). In that scenario, one
can neglect the ionic contribution, and consequently, $E_{B}$ remains
unchanged\citep{chapter1-65,chapter1-66}. From Table \ref{tab:4}
and Table \ref{tab:6}, it is found that $E_{B}\gg\hbar\omega_{LO}$
for ASnX$_{3}$ CPs; therefore, ionic screening to the dielectric
function can be disregarded. The $E_{B}$ values obtained from standard
first-principles GW-BSE calculations are found in the range of 0.05$-$0.24
eV, which are comparable to that of transition metal-based traditional
CPs (0.02$-$0.26 eV)\citep{chapter3-19,chapter1-63,chapter5-16}
but a little higher than those observed in conventional lead-based
HPs (0.01$-$0.10 eV)\citep{chapter5-11,chapter5-12,chapter5-14}
(for details, see Table S14 of the Supplemental Material). The $E_{B}$
values obtained from our GW-BSE calculations indicate relatively strong
excitonic effects in these materials. While this range may appear
too high for direct photovoltaic applications, it is worth noting
that these values represent intrinsic properties of isolated bulk
materials. In practical device architectures, factors such as dielectric
screening from surrounding layers, nanostructuring, or heterostructures
can significantly reduce the effective $E_{B}$, potentially bringing
it within the desired range for photovoltaic applications. Furthermore,
these materials may be more suited for other optoelectronic applications,
such as light-emitting diodes, photodetectors, or excitonic solar
cells, where stronger excitonic effects are advantageous. Here, CaSnSe$_{3}$
exhibits the lowest $E_{B}$ value, attributed to its higher electronic
dielectric constant ($\varepsilon_{\infty}$), which reduces the Coulomb
interaction between the electron and hole.

\begin{table}[H]
\caption{\label{tab:4}Calculated excitonic parameters for chalcogenide perovskites.
Here, $E_{g}^{d}$ represents the direct G$_{0}$W$_{0}$@PBE bandgap,
$E_{exc}$ represents the optical bandgap, $E_{B}$ represents the
exciton binding energy, $T_{exc}$ represents the excitonic temperature,
$r_{exc}$ is the exciton radius, $|\phi_{n}(0)|^{2}$ is the probability
of a wave function at zero separation, $\varepsilon_{\infty}$ is
the electronic dielectric constant, $\varepsilon_{static}$ is the
static dielectric constant, $E_{Bu}$ and $E_{Bl}$ are the upper
and lower bounds of the exciton binding energy, respectively.}

\centering{}{\small{}}%
\begin{tabular}{cccccccccccc}
\hline 
\multirow{3}{*}{{\small{}Configurations}} & \multicolumn{6}{c}{{\small{}First-principles method (GW-BSE)}} &  & \multicolumn{4}{c}{{\small{}Wannier\textminus Mott model}}\tabularnewline
\cline{2-7} \cline{3-7} \cline{4-7} \cline{5-7} \cline{6-7} \cline{7-7} \cline{9-12} \cline{10-12} \cline{11-12} \cline{12-12} 
 & {\small{}$E_{g}^{d}$} & {\small{}$E_{o}$} & {\small{}$E_{B}$} & {\small{}$T_{exc}$} & {\small{}$r_{exc}$} & {\small{}$|\phi_{n}(0)|^{2}$} &  & \multirow{2}{*}{{\small{}$\varepsilon_{\infty}$}} & {\small{}$E_{Bu}$} & \multirow{2}{*}{{\small{}$\varepsilon_{static}$}} & {\small{}$E_{Bl}$}\tabularnewline
 & {\small{}(eV)} & {\small{}(eV)} & {\small{}(eV)} & {\small{}(K)} & {\small{}(nm)} & {\small{}($10^{26}$$m^{-3}$)} &  &  & {\small{}(meV)} &  & {\small{}(meV)}\tabularnewline
\hline 
{\small{}CaSnS$_{3}$} & {\small{}1.44} & {\small{}1.23} & {\small{}0.21} & {\small{}2435} & {\small{}1.72} & {\small{}0.63} &  & {\small{}7.47} & {\small{}56.06} & {\small{}34.70} & {\small{}2.60}\tabularnewline
{\small{}SrSnS$_{3}$} & {\small{}1.50} & {\small{}1.26} & {\small{}0.24} & {\small{}2783} & {\small{}1.28} & {\small{}1.52} &  & {\small{}7.72} & {\small{}72.79} & {\small{}32.55} & {\small{}4.09}\tabularnewline
{\small{}BaSnS$_{3}$} & {\small{}1.28} & {\small{}1.12} & {\small{}0.16} & {\small{}1855} & {\small{}1.13} & {\small{}2.21} &  & {\small{}8.97} & {\small{}71.16} & {\small{}45.61} & {\small{}2.75}\tabularnewline
{\small{}CaSnSe$_{3}$} & {\small{}0.79} & {\small{}0.74} & {\small{}0.05} & {\small{}580} & {\small{}9.21} & {\small{}0.004} &  & {\small{}15.15} & {\small{}5.16} & {\small{}70.08} & {\small{}0.24}\tabularnewline
{\small{}SrSnSe$_{3}$} & {\small{}0.88} & {\small{}0.78} & {\small{}0.10} & {\small{}1159} & {\small{}4.79} & {\small{}0.03} &  & {\small{}13.39} & {\small{}11.23} & {\small{}56.98} & {\small{}0.62}\tabularnewline
{\small{}BaSnSe$_{3}$} & {\small{}1.04} & {\small{}0.91} & {\small{}0.13} & {\small{}1507} & {\small{}1.48} & {\small{}0.99} &  & {\small{}12.67} & {\small{}38.46} & {\small{}44.66} & {\small{}3.10}\tabularnewline
\hline 
\end{tabular}{\small\par}
\end{table}

In addition, we have validated the $E_{B}$ using the Wannier-Mott
method\citep{chapter1-63,chapter2-38} through the Eq. \ref{eq:1}.
This method helps establish a qualitative understanding of excitonic
properties and their dependence on effective masses and dielectric
screening, serving as a sanity check for trends observed in the ab
initio BSE calculations. The Wannier-Mott model is traditionally more
applicable to direct bandgap materials. However, for indirect bandgap
materials in our study, the effective masses and, subsequently, the
reduced masses are calculated at their direct band edges (for details,
see table \ref{tab:3}). Here, $\mathrm{\varepsilon_{eff}}$ lies
in between optical ($\varepsilon_{\infty}$) and static ($\varepsilon_{static}$)
dielectric constant. The optical or electronic dielectric constants
($\varepsilon_{\infty}$) at the zero frequency limit are obtained
using the BSE method. Subsequently, the DFPT calculations are performed
to compute the ionic contribution to the dielectric function ($\varepsilon_{ion}$),
and therefore, the static dielectric constant is calculated as, $\varepsilon_{static}$
= ($\varepsilon_{\infty}$ + $\varepsilon_{ion}$). In our study,
the upper ($E_{Bu}$) and lower ($E_{Bl}$) bounds of exciton binding
energy are estimated based on the $\varepsilon_{\infty}$ and $\varepsilon_{static}$,
respectively (see Table \ref{tab:4}). The upper bound values are
smaller but align more closely with the $E_{B}$ calculated using
the GW-BSE method than the lower bound values; however, the overall
trend remains consistent. This suggests that in chalcogenide perovskites,
the electronic contribution to dielectric screening is more prominent
than the ionic contribution.

Furthermore, additional investigations into the ionic (phonon) contribution
to the exciton binding energy ($E_{B}$) are conducted for the examined
CPs. This is because the standard BSE approach within an ab initio
framework captures only static screening from electrons to calculate
the exciton binding energy. However, it's noteworthy that dynamic
electron-electron interactions or electron-phonon coupling may play
a crucial role in certain materials, particularly those with significant
electron-phonon interactions or where phonons significantly influence
optoelectronic properties. Filip and collaborators\citep{chapter3-39}
have recently broadened the discussion to incorporate phonon screening
into the analysis of exciton binding energy using equation \ref{eq:2}.
They achieved this by considering four specific material parameters:
reduced effective mass, static and optical dielectric constants, and
the frequency of the longitudinal optical phonon mode ($\omega_{LO}$),
while assuming isotropic and parabolic electronic band dispersion.
Table \ref{tab:5} indicates that phonon screening leads to a reduction
in the $E_{B}$ ranging from 5.58\% to 17.87\%, with the modified
$E_{B}$ values ranging from 0.04 to 0.23 eV. While phonon screening
does decrease the exciton binding energy, the reduction is not particularly
substantial, except in the case of CaSnSe$_{3}$. This also suggests
that in most chalcogenide perovskites, the electronic contribution
to dielectric screening outweighs the ionic (or phonon) contribution.

\begin{table}[H]
\caption{\label{tab:5}Calculated exciton binding energy ($E_{B}$), phonon
screening corrections ($\Delta E_{B}^{ph}$), percentage of phonon
screening contribution to the reduction of exciton binding energy
(\%), and corrected values of exciton binding energy $(E_{B}+\Delta E_{B}^{ph})$
for chalcogenide perovskites.}

\centering{}%
\begin{tabular}{ccccc}
\hline 
Configurations & $E_{B}$ (meV) & $\Delta E_{B}^{ph}$ (meV) & Reduction of $E_{B}$ (\%) & $(E_{B}+\Delta E_{B}^{ph})$ (meV)\tabularnewline
\hline 
CaSnS$_{3}$ & 210 & -14.68 & 6.99 & 195.32\tabularnewline
SrSnS$_{3}$ & 240 & -13.39 & 5.58 & 226.61\tabularnewline
BaSnS$_{3}$ & 160 & -12.42 & 7.76 & 147.58\tabularnewline
CaSnSe$_{3}$ & 50 & -8.93 & 17.87 & 41.07\tabularnewline
SrSnSe$_{3}$ & 100 & -8.17 & 8.17 & 91.83\tabularnewline
BaSnSe$_{3}$ & 130 & -8.08 & 6.22 & 121.92\tabularnewline
\hline 
\end{tabular}
\end{table}

Several excitonic parameters, including excitonic temperature ($T_{exc}$),
exciton radius ($r_{exc}$), and probability of wave function ($|\phi_{n}(0)|^{2}$)
for $e-h$ pair at zero separation are also computed using the above
quantities ($E_{B}$, $\varepsilon_{\infty}$, and $\mu^{*}$) to
have a definitive estimation of the excitonic properties (for details,
see the Supplemental Material). Based on the analysis of exciton radius
($r_{exc}$), it is observed that in all investigated materials, the
electron-hole pairs are distributed over multiple lattice constants,
supporting the applicability of the Wannier-Mott model. The inverse
of $|\phi_{n}(0)|^{2}$ can be used to qualitatively characterize
the exciton lifetime ($\tau_{exc}$), which is listed in Table \ref{tab:4}
(for details, see the Supplemental Material). Consequently, the $\tau_{exc}$
values for the investigated CPs are in the order CaSnSe$_{3}$ $>$
SrSnSe$_{3}$ $>$ CaSnS$_{3}$ $>$ BaSnSe$_{3}$ $>$ SrSnS$_{3}$
$>$ BaSnS$_{3}$. For efficient photo-conversion in solar cells,
the exciton lifetime ($\tau_{exc}$) should be sufficiently high to
extract photo-generated charge carriers before recombination occurs.
A longer exciton lifetime corresponds to a lower carrier recombination
rate, which enhances the quantum yield and conversion efficiency.
Overall, these properties significantly enhance the efficiency of
ASnX$_{3}$ CPs, making them promising for potential optoelectronic
applications.

\subsection{Polaronic Properties:}

In polar semiconductors, such as halide perovskite and its derivatives,
the scattering mechanism near room temperature is dominant due to
the interaction between charge carriers and the macroscopic electric
field produced by longitudinal optical phonon (LO)\citep{chapter2-13}.
This interaction strongly influences charge carrier mobility of the
system and is anticipated to be the same for the materials of our
interest\citep{chapter3-19,chapter1-63}. Therefore, the Fr\"ohlich
mesoscopic model\citep{chapter2-13,chapter2-20,chapter2-51} is used
to describe this interaction and defined by the dimensionless Fr\"ohlich
parameter $\alpha$ using Eq. \ref{eq:3}. The calculated values of
$\alpha$ related to Fr\"ohlich interaction for electrons and holes
are given in Table \ref{tab:6} and Table S12, respectively. Strong
electron (hole)-phonon coupling is indicated by $\alpha>10$, while
$\alpha\ll1$ often suggests weak coupling\citep{chapter2-20}. Our
results show that the value of $\alpha$ for ASnX$_{3}$ CPs lies
in the range of 0.64$-$2.92, suggesting weak to intermediate electron
(hole)-phonon coupling for the investigated systems. Additionally,
it has been observed that Sn-based CPs exhibit less pronounced polaronic
effects compared to transition metal-based CPs\citep{chapter3-19,chapter1-63,chapter5-16}.
This weaker electron-phonon coupling can be attributed to a lower
electron effective mass and a higher electronic dielectric constant,
while stronger coupling in transition metal-based CPs results from
the opposite factors. Here, CaSnSe$_{3}$ exhibits smaller electron-phonon
coupling, while BaSnS$_{3}$ demonstrates larger electron-phonon coupling
for the same reasons. In this context, the specific free volume of
these CPs is also evaluated to obtain a qualitative knowledge of the
strength of electron-phonon coupling (for details, see the Supplemental
Material).

Notably, polaron formation can lead to a decrease in the electron
and hole QP energies. This polaron energy ($E_{p}$) can also be calculated
using $\alpha$ by using Eq. \ref{eq:4}. The QP gap, which results
from the polaron energy for electrons and holes, is reduced by 76.37,
81.16, 85.32, 18.58, 25.37, and 43.65 meV for CaSnS$_{3}$, SrSnS$_{3}$,
BaSnS$_{3}$, CaSnSe$_{3}$, SrSnSe$_{3}$, and BaSnSe$_{3}$, respectively.
Comparing the values of $E_{B}$ from Table \ref{tab:4} with the
QP gap, it is evident that the energy of charge-separated polaronic
states is lower than that of the bound exciton states. This suggests
that the charge-separated polaronic states are less stable than the
bound excitons.

The other parameters for the polarons, i.e., polaron mass ($m_{p}$)
and polaron mobility ($\mu_{p}$), are also estimated using Eq. \ref{eq:5}
and Eq. \ref{eq:6}, respectively. These parameter values are listed
in Table \ref{tab:6} and Table S12 for electrons and holes, respectively.
One can confirm the enhanced carrier-lattice interactions by higher
$m_{p}$ values, and for this case, the charge carrier mobility decreases
than the non-polar or less polar compounds. For example, Ca- and Sr-based
compounds exhibit weaker carrier (electron)-lattice interactions compared
to Ba-based compounds, resulting in higher mobility for Ca- and Sr-based
materials. Also, CaSnSe$_{3}$ CP exhibits the highest polaron mobility
for electrons, which is obvious due to weak electron-phonon coupling
($\alpha$ = 0.64). Overall, these post-transition metal Sn-based
CPs are shown to have ambipolar properties and much-improved polaron
mobility for electrons (21.33$-$416.02 cm$^{2}$V$^{-1}$s$^{-1}$)
and holes (7.02$-$260.69 cm$^{2}$V$^{-1}$s$^{-1}$, see Table S12
of the Supplemental Material) than the conventional lead-based HPs
(57$-$290 cm$^{2}$V$^{-1}$s$^{-1}$ for electrons and 97$-$230
cm$^{2}$V$^{-1}$s$^{-1}$ for holes, respectively)\citep{chapter2-20,chapter5-15}
as well as transition metal (Zr and Hf) based traditional CPs (6.84$-$77.59
cm$^{2}$V$^{-1}$s$^{-1}$ for electrons and 3.76$-$100.49 cm$^{2}$V$^{-1}$s$^{-1}$
for holes, respectively)\citep{chapter3-19,chapter1-63,chapter5-16}.
The higher polaron mobility in Sn-based CPs is attributed to less
prominent polaronic effects than conventional lead-based HPs and transition
metal-based CPs.

\begin{table}[H]
\caption{\label{tab:6}Calculated polaron parameters for electrons in chalcogenide
perovskites. Here, $\omega_{LO}$ represents the characteristic phonon
angular frequency, $\theta_{D}$ represents the Debye temperature,
$\alpha$ represents the Fr\"ohlich interaction parameter, $E_{p}$
represents the polaron energy, $m_{p}$ represents the effective mass
of the polaron, and $\mu_{p}$ represents the polaron mobility, respectively.}

\centering{}%
\begin{tabular}{ccccccc}
\hline 
\multicolumn{1}{c}{Configurations} & $\omega_{LO}$ (THz) & $\theta_{D}$ (K) & $\alpha$ & $E_{p}$ (meV) & $m_{p}/m^{*}$ & $\mu_{p}$ (cm$^{2}$V$^{-1}$s$^{-1}$)\tabularnewline
\hline 
CaSnS$_{3}$ & 4.78 & 229.51 & 1.75 & 35.39 & 1.37 & 42.29\tabularnewline
SrSnS$_{3}$ & 4.44 & 213.18 & 2.10 & 39.61 & 1.46 & 22.64\tabularnewline
BaSnS$_{3}$ & 3.97 & 190.62 & 2.11 & 35.59 & 1.46 & 21.33\tabularnewline
CaSnSe$_{3}$ & 3.20 & 153.65 & 0.64 & 8.55 & 1.12 & 416.02\tabularnewline
SrSnSe$_{3}$ & 2.76 & 132.52 & 0.97 & 11.22 & 1.18 & 173.92\tabularnewline
BaSnSe$_{3}$ & 2.88 & 138.28 & 1.62 & 19.71 & 1.33 & 32.86\tabularnewline
\hline 
\end{tabular}
\end{table}

\subsection{Spectroscopic Limited Maximum Efficiency:}

The above-discussed properties indicate that the investigated systems
hold great potential for photovoltaic applications, and to validate
this, we also calculated their power conversion efficiency (PCE).
The theoretical PCE of each system is calculated using the spectroscopic
limited maximum efficiency (SLME) method. SLME was introduced by Yu
and Zunger\citep{chapter3-30} (for details, see the Supplemental
Material), which is an improved version of the Shockley-Queisser (SQ)
limit\citep{chapter3-31}. The latter is less realistic since it disregards
the losses resulting from radiative recombinations due to the non-conservation
of the absorbed photon momentum. SLME incorporates the magnitude of
the bandgap and its nature (direct or indirect), the shape of absorption
spectra, the thickness of the absorber layer, the material-dependent
non-radiative recombination losses, and the temperature. The standard
solar spectrum (AM-1.5G), the absorption coefficient, thickness, and
the electronic G$_{0}$W$_{0}$@PBE bandgap are thus used as inputs
to evaluate the theoretical SLME of ASnX$_{3}$ (A = Ca, Sr, Ba; X
= S, Se) CPs at 293.15 K temperature.

\begin{figure}[H]
\begin{centering}
\includegraphics[width=1\textwidth,height=1\textheight,keepaspectratio]{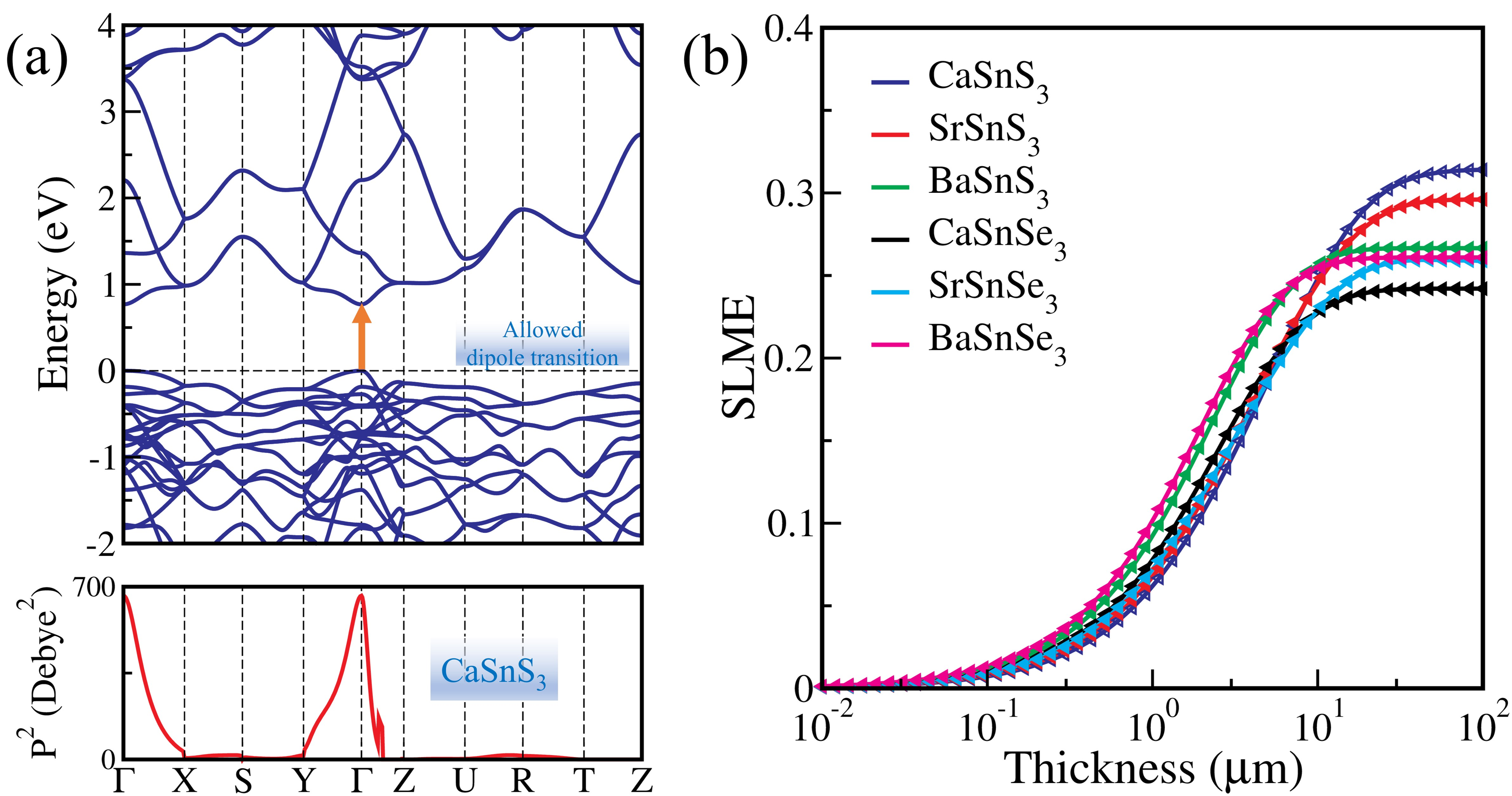}
\par\end{centering}
\caption{\label{fig:4}(a) Electronic band structure and transition probability
(square of the transition dipole moment matrix elements) of CaSnS$_{3}$
calculated using PBE, and (b) spectroscopic limited maximum efficiency
(SLME) of ASnX$_{3}$ (A = Ca, Sr, Ba; X = S, Se) chalcogenide perovskites
calculated using BSE@G$_{0}$W$_{0}$@PBE method.}
\end{figure}

In addition, the optical transition possibility from VBM to CBM for
these CPs has been confirmed through the computation of transition
dipole moment matrix (P) elements; its square (P$^{2}$) gives the
transition probability between the initial (VBM) and the final (CBM)
state. Figure \ref{fig:4}(a) shows the optically allowed dipole transition
at $\Gamma$-point for CaSnS$_{3}$, and for other configurations,
see Figure S7. Despite having an indirect electronic bandgap, BaSnS$_{3}$
and BaSnSe$_{3}$ exhibit an optically allowed dipole transition at
their direct band edge. However, when $E_{g}^{da}$ is not the minimum
bandgap of the materials (i.e., $E_{g}$ $\neq$ $E_{g}^{da}$), non-radiative
recombination plays a vital role in the SLME calculation and the radiative
recombinations vary with a factor $f_{r}=e^{-\Delta/k_{B}T}$, where
$\Delta=\text{\ensuremath{E_{g}^{da}} }-\text{\ensuremath{E_{g}}}$,
$k_{B}$ is the Boltzmann constant, and T is the temperature\citep{chapter3-30}
(for details, see the Supplemental Material).

Next, the thickness dependence of SLME has been computed for all the
ASnX$_{3}$ (A = Ca, Sr, Ba; X = S, Se) compounds using BSE@G$_{0}$W$_{0}$@PBE
method and plotted in Figure \ref{fig:4}(b). It is clear that the
SLME rises as thickness increases and eventually saturates beyond
a certain thickness. The maximum SLME is found to be 31.20\% for CaSnS$_{3}$,
which is consistent with the previously documented theoretical efficiency
of 32.45\% at 10 $\mu$m thickness\citep{chapter3-19} and is higher
than CH$_{3}$NH$_{3}$PbI$_{3}$ (28.97\% at 2 $\mu$m)\citep{chapter5-13}.
The highest SLME values for SrSnS$_{3}$, BaSnS$_{3}$, CaSnSe$_{3}$,
SrSnSe$_{3}$, and BaSnSe$_{3}$ compounds are calculated as 29.53\%,
26.66\%, 24.20\%, 25.95\%, and 26.09\%, respectively. These values
of SLME are favorable for the photovoltaic applications compared to
transition metal-based traditional CPs and other halide perovskites\citep{chapter3-19,chapter3-33,chapter3-34,chapter3-32,chapter1-63,chapter5-16}
(for details, see Table S14 of the Supplemental Material).

\subsection{Analysis of SCAPS-1D results:}

\begin{figure}[H]
\begin{centering}
\includegraphics[width=1\textwidth,height=1\textheight,keepaspectratio]{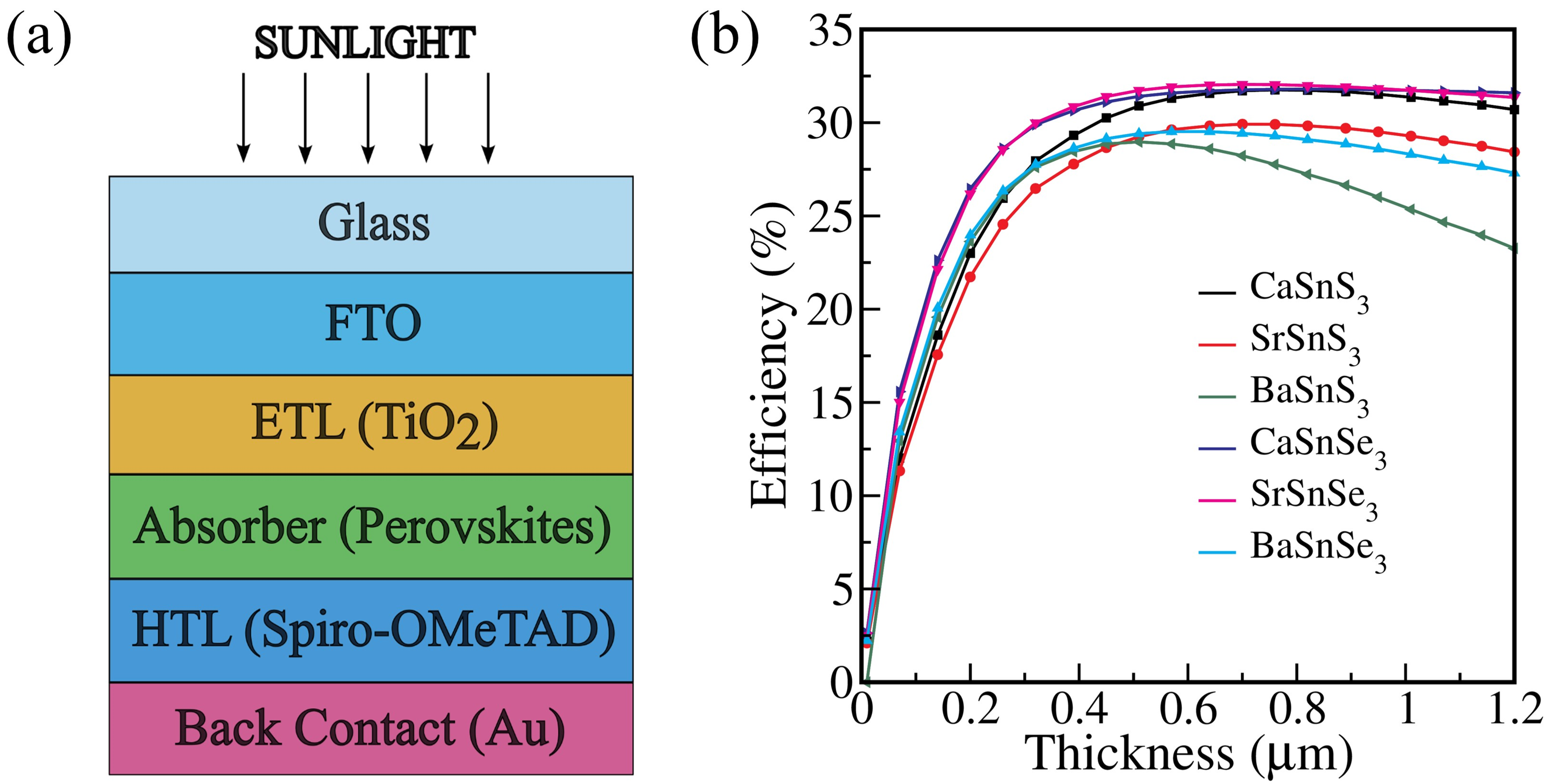}
\par\end{centering}
\caption{\label{fig:5}(a) Schematic of the simulated device structure and
(b) power conversion efficiency (PCE) of the ASnX$_{3}$ (A = Ca,
Sr, Ba; X = S, Se) chalcogenide perovskites-based PSC using SCAPS-1D.}
\end{figure}

To verify the photovoltaic potential of the investigated systems and
our calculated SLME, we also conducted a detailed analysis using the
Solar Cell Capacitance Simulator in 1 Dimension (SCAPS-1D). This simulation
allowed us to evaluate key performance metrics and gain deeper insights
into the efficiency and operational behavior of the devices. Here,
the performance of PSC is evaluated by varying the thickness of the
absorber layer while keeping the ETL, HTL, and FTO thicknesses constant.
The absorber layer thickness is varied from 0.01 $\mu$m to 1.2 $\mu$m,
and it found that the thickness of the perovskite layer has a significant
impact on the performance of the PSC {[}see Figure \ref{fig:5}(b){]}.
After reaching the optimum efficiency, the effect of the thickness
of the perovskite layer seems to saturate, which agrees with our SLME
calculations. The obtained results are summarized in the Table \ref{tab:7}.
Our findings indicate that the device simulation results are consistent
with the SLME work, specifically for ASnS$_{3}$ (A = Ca, Sr, Ba).
On the other hand, the minor discrepancies between the device simulation
results for ASnSe$_{3}$ (A = Ca, Sr, Ba) and the SLME findings may
be attributed to the materials used in the ETL, HTL, and FTO layers.
In our study, SLME provides an idealized efficiency limit based solely
on the material's intrinsic optical properties, with its behavior
improving for thicker films as absorption approaches a step function,
aligning with the SQ limit. However, SCAPS-1D accounts for additional
real-world effects, such as carrier transport, recombination losses,
and device architecture, which can result in a performance decay for
thicker films beyond 1 $\mu$m due to increased recombination or reduced
carrier collection efficiency. This distinction highlights the necessity
of combining both methods to achieve a comprehensive assessment of
the material's optoelectronic potential, effectively bridging the
gap between idealized theoretical predictions and practical device
limitations. In general, SCAPS-1D proves to be a highly reliable software,
as its simulation results confirm the high efficiency of the investigated
materials, closely aligning with theoretical predictions.

\begin{table}[H]
\caption{\label{tab:7}Photovoltaic parameters of ASnX$_{3}$ (A = Ca, Sr,
Ba; X = S, Se) chalcogenide perovskites based solar cells using SCAPS-1D.}

\centering{}{\small{}}%
\begin{tabular}{ccccc}
\hline 
{\small{}Optimized Device} & {\small{}$V_{oc}$ (V)} & {\small{}$J_{sc}$ (mA/cm$^{2}$)} & {\small{}FF (\%)} & {\small{}PCE (\%)}\tabularnewline
\hline 
{\small{}FTO/TiO$_{2}$/CaSnS$_{3}$/Spiro-OMeTAD/Au} & {\small{}1.2885} & {\small{}27.49796} & {\small{}89.63} & {\small{}31.76}\tabularnewline
{\small{}FTO/TiO$_{2}$/SrSnS$_{3}$/Spiro-OMeTAD/Au} & {\small{}1.3239} & {\small{}25.06713} & {\small{}90.15} & {\small{}29.92}\tabularnewline
{\small{}FTO/TiO$_{2}$/BaSnS$_{3}$/Spiro-OMeTAD/Au} & {\small{}0.9871} & {\small{}33.93041} & {\small{}86.49} & {\small{}28.97}\tabularnewline
{\small{}FTO/TiO$_{2}$/CaSnSe$_{3}$/Spiro-OMeTAD/Au} & {\small{}0.7255} & {\small{}55.85582} & {\small{}82.92} & {\small{}31.80}\tabularnewline
{\small{}FTO/TiO$_{2}$/SrSnSe$_{3}$/Spiro-OMeTAD/Au} & {\small{}0.7750} & {\small{}49.34989} & {\small{}83.80} & {\small{}32.05}\tabularnewline
{\small{}FTO/TiO$_{2}$/BaSnSe$_{3}$/Spiro-OMeTAD/Au} & {\small{}0.8050} & {\small{}43.53916} & {\small{}84.22} & {\small{}29.52}\tabularnewline
\hline 
\end{tabular}{\small\par}
\end{table}

Overall, the study of all the above-discussed properties collectively
shows that Se-based CaSnSe$_{3}$ and SrSnSe$_{3}$ demonstrate lower
exciton binding energies and higher charge carrier mobilities than
the rest of the investigated systems. However, their SLME is also
relatively low. On the other hand, S-based CaSnS$_{3}$ and SrSnS$_{3}$
perovskites exhibit higher SLME despite having higher exciton binding
energies and lower charge carrier mobilities compared to the other
examined materials. This improved SLME is primarily attributed to
the optimal bandgaps of CaSnS$_{3}$ and SrSnS$_{3}$ compared to
their selenium (Se) counterparts, specifically from the perspective
of solar cells. This reveals a trade-off between exciton binding energy,
charge carrier mobility, and the bandgap of these materials. This
suggests that integrating a combination of sulfur (S) and selenium
(Se) atoms in these materials could be advantageous for potential
application in solar cells, meriting further investigation.

\section{Conclusions:}

In conclusion, we have carried out a comprehensive study to investigate
the ground- and excited-state properties of post-transition metal
Sn-based distorted chalcogenide perovskites (ASnX$_{3}$; A = Ca,
Sr, Ba, and X = S, Se) under the framework of state-of-the-art DFT
combined with DFPT and MBPT (viz., GW and BSE). The mechanical properties
confirm the stability, and Pugh\textquoteright s and Poisson\textquoteright s
ratios reveal the ductile nature of these perovskites. The bandgaps
calculated using G$_{0}$W$_{0}$@PBE method are in the range of 0.79$-$1.50
eV, and the small effective masses of electrons suggest good charge
carrier mobility, which is advantageous for energy-harvesting properties.
Furthermore, they have a high optical absorption coefficient ($>$
10$^{4}$ cm$^{-1}$) and an optical bandgap extending from the near-infrared
to the visible range, which is indicated by the BSE calculations.
The exciton binding energies of these compounds, ranging from 0.04
to 0.23 eV, are similar to those of transition metal (Zr and Hf) based
CPs. Also, these perovskites have smaller carrier-phonon coupling
strengths than the conventional lead-based HPs as well as transition
metal-based CPs, which leads to better polaron mobility for electrons
(21.33$-$416.02 cm$^{2}$V$^{-1}$s$^{-1}$) and holes (7.02$-$260.69
cm$^{2}$V$^{-1}$s$^{-1}$). Additionally, the Fr\"ohlich mesoscopic
model implies that the charge-separated polaronic states have lower
stability in comparison to bound excitons. Lastly, the SLME method
forecasts that one could achieve the highest PCE of up to 31.20\%
by employing them, which is corroborated by conventional device (FTO/TiO$_{2}$/ASnX$_{3}$/Spiro-OMeTAD/Au)
simulations using SCAPS-1D software. The enhanced properties of Sn-based
CPs can be attributed to the distinct valence state of Sn in comparison
to transition metal-based CPs. Overall, these results are anticipated
to expedite the research and use of chalcogenide perovskites in optoelectronic
applications, in general, and solar cell technology, in particular.
\begin{acknowledgments}
The authors would like to acknowledge the Council of Scientific and
Industrial Research (CSIR), Government of India {[}Grant No. 3WS(007)/2023-24/EMR-II/ASPIRE{]}
for financial support. S.A. would also like to acknowledge the CSIR,
India {[}Grant No. 09/1128(11453)/2021-EMR-I{]} for the Senior Research
Fellowship. S. D. thanks Prof. Marc Burgelman, Department of Electronics
and Information Systems (ELIS), University of Gent, for providing
the SCAPS-1D software. The authors acknowledge the High Performance
Computing Cluster (HPCC) 'Magus' at Shiv Nadar Institution of Eminence
for providing computational resources that have contributed to the
research results reported within this paper.
\end{acknowledgments}

\bibliographystyle{apsrev4-2}
\bibliography{refs}

\end{document}